\documentclass{article}

\usepackage{graphicx}
\usepackage{authblk}

\title{Inequalities faced by women in access to permanent positions in astronomy in France}

\author{Olivier Bern\'e}
\author{Alexia Hilaire}
\affil[1]{Institut de Recherche en Astrophysique et Planetologie, Universit\'e de Toulouse, CNRS, CNES, UPS, Toulouse, France}

\begin{document}

\maketitle



We investigate inequalities in access 
 to permanent positions in professional astronomy in France, focusing on the hiring stage.
 We use results from a national survey conducted 
 on behalf of the French society of astronomy and astrophysics (SF2A) aimed at 
 young astronomers holding a PhD obtained in France, and answered by over 300 
 researchers.

 We find that women are nearly two times less likely than 
 men to be selected by the (national or local) committees attributing permanent positions ($p=0.06$). 
 We also find that applicants who did their undergraduate studies in an elite school (``Grande \'Ecole''), 
 where women are largely under-represented, rather than in a university, 
 are nearly three times more likely to succeed in obtaining a position ($p=0.0026$).
 Our analysis suggests the existence of two biases in committees attributing 
 permanent positions in astronomy in France: a gender bias, and a form of elitism. 
 These biases against women in their professional life 
 impacts their personal life as our survey shows 
 that a larger fraction of them declare that having 
 children can have a negative effect on their careers. 
 They are half as many as men having children in the
 sample. National committees (such as the CNRS) have acknowledged 
 this issue for several years now, hence one can hope that 
 changes will be seen in the next decade.


\section*{Introduction}

Gender inequalities in science have been the focus of numerous studies 
involving sociology, psychology, economics, etc. Although these inequalities have been 
well identified for several decades, little progress has been achieved \cite{val99}, 
in particular in several  fields of Science Technology Engineering 
and Mathematics (STEM) where women remain under-represented \cite{bee11}, 
including in physics \cite{glo00}. Astronomy is also a field where women are 
under-represented. For instance, there are only 18\% of members of the international 
astronomical union (IAU, the largest organisation of professional astronomers) who are 
females. 
This under-representation is often attributed to some forms of discrimination against women
astronomers in their careers, one striking example in the discipline being 
access to observing time on telescopes. Decades ago already it was reported by e.g. Vera Rubin, 
that access to Mount Wilson observatory for women was a struggle \cite{rub81}. 
Recent empirical studies have shown that gender biases in the attribution of telescope 
time persist at major facilities such as the Atacama Millimeter Array \cite{car19},
 the Hubble Space Telescope \cite{rei14}, the European Southern Observatory \cite{pat16},
 and the NRAO \cite{lon16}. Although women are now better represented 
in large scientific collaborations in astronomy, only few of them serve as leaders in 
these teams  \cite{luc17}, and when it gets to receiving a share of a prize, 
they can be forgotten \cite{ber18}. 
Regarding the success of women in obtaining permanent/faculty positions, 
a study by Flaherty \cite{flah18} has shown for instance that female astronomers 
leave the academic labour market at a much higher rate then men. This result was 
however mitigated by another study in which it was found that the rate at which 
women are hired on long-term positions in astronomy (or closely related fields) 
in the United States is similar to that of men \cite{per19}. The existence 
of gender inequalities in the field of astronomy has been covered mostly in the 
United States, but interest has sparked in other countries (see e.g. \cite{bro11,mat14, sed18}). 
Studies for the case of France, which is one important nation in the field, and where 
only 22\% of astronomers holding a permanent position are women lack. In addition, while there are, 
as we have seen, several studies on gender biases in astronomy, research addressing 
the articulation between gender and other types of biases at hiring, as well as the impact of 
such biases on the personal life of female astronomers are scarce. 
Here we provide an example of such an analysis for the case of France using 
the results of a survey conducted in 2017 amongst young astronomers in this country.

\section{A national survey on French astronomy}

The survey was elaborated by the council of the French Society of Astronomy and 
Astrophysics (SF2A, www.sf2a.eu), and conducted using an online form. It was 
aimed at researchers who obtained their PhD in astronomy between 2007 and 2017, in France,
but not limited to French nationals.  
Questions in the survey concerned the profile of the young astronomers,
their current status in particular in terms of employment, their applications to
permanent positions and the perception  of their careers. Profile includes gender, 
year of PhD, city where PhD was obtained and type of undergraduate studies. 
Status concerns their current position, either in or outside of research, public or private, 
permanent or short-term as well as a personal status i.e. whether they have children
or not. A fraction of the survey was dedicated to applications 
by young astronomers to permanent positions in academia, so as to identify if 
they had or not applied to permanent positions, of which type (researcher, 
dep. astronomer, ass. prof), and whether they had succeeded or not.  
Several questions were devoted to their perception on their careers,
in particular if they ``[...] believe that having children can be an obstacle in [their] career[s]?''. 
The survey was distributed through national professional astronomy canals, including the 
newsletter of SF2A. 


In total, the survey received 301 answers. This is approximately a third
of the total population: about 1000 students obtained their PhD in astronomy 
between 2007 and  2017 in France according to data collected by 
the doctoral school of the Paris Observatory. Amongst the 301 answers, 
198 self declare as men, 97 as women, and 6 as "do not want to answer". 
This corresponds to 33\% women amongst those who answered, which is very close to 
the ratio that is observed, independently, of women who defended a PhD in astronomy 
in France between 
2012 and 2018 (32\% 
of PhDs in astronomy in France were defended by women, according to data collected 
by the doctoral school of the Paris Observatory).


\begin{figure}
\centerline{\includegraphics[height = 8cm, keepaspectratio]{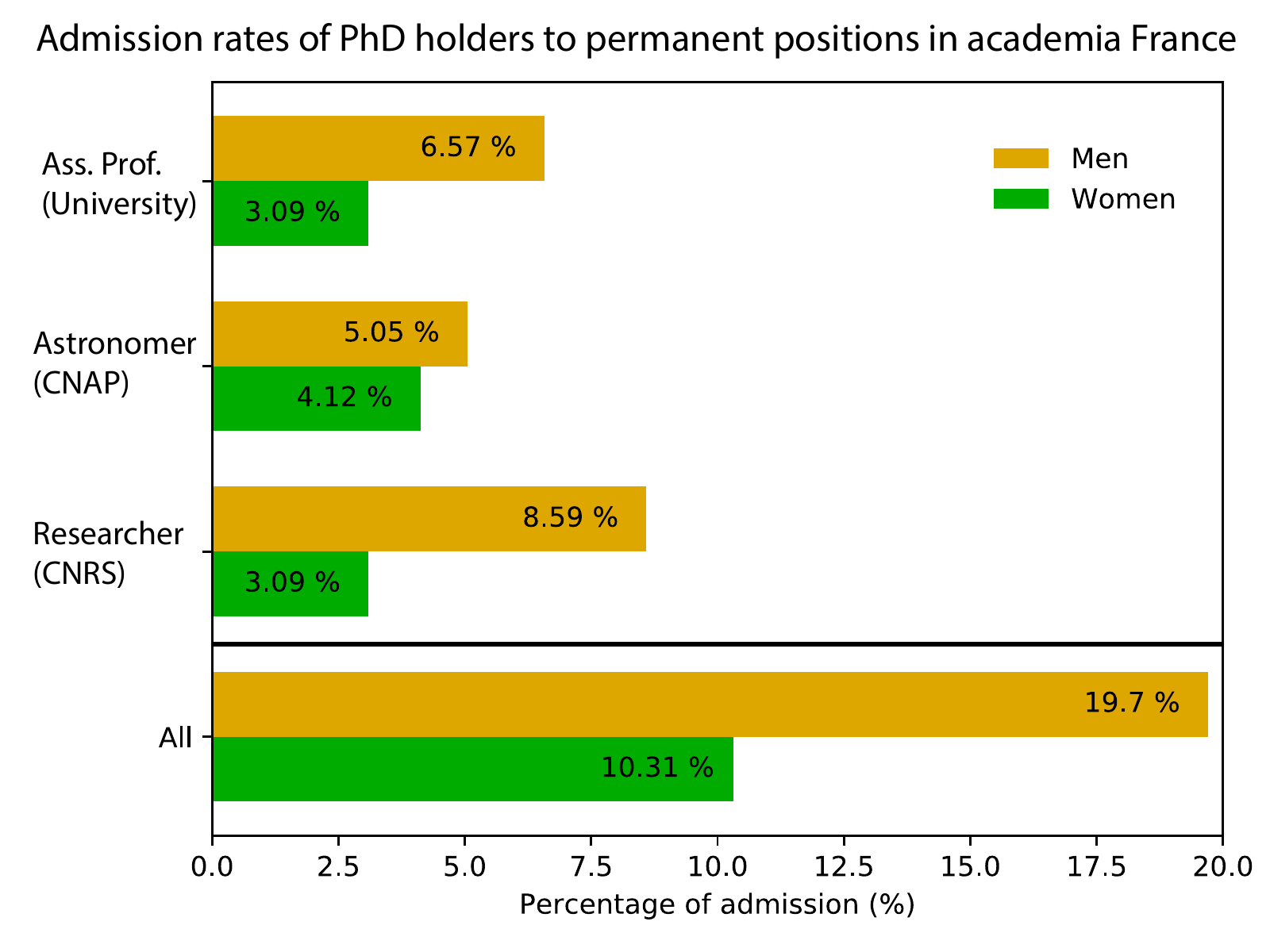}}
\caption{Admission rates of applicants ($<10$ years after PhD)) to permanent positions in France. 
\label{Fig_admission-rate}}
\end{figure}

\section{Elitism and gender biases in French astronomy}

We studied the admission rates to permanent positions, defined as
the ratio between the number people who have applied, to the number of applicants 
who succeeded in obtaining a position for the three major types of permanent positions
available in astronomy in France (see box for details about types of permanent positions 
in France).  It is important to note that hiring at a permanent position in France mostly occurs less then 
10 year after PhD. Hence the measured admission rate, which is 16.6\%, can be considered 
as an estimate of the total rate at which the academic system 
absorbs applicants in astronomy. 

We first investigate the admission rates as a function of academic background. In the French (mostly public) education system 
two systems cohabit in science:  universities, which are non-selective and where most students 
go to, and elite schools i.e. ``Grandes \'Ecoles'', which are selective and where the ``best'' 
high school students go to. The two most prestigious and selective ``Grandes \'Ecoles'' are the 
\'Ecole Normale Supérieure (ENS), and the \'Ecole Polytechnique, which can be compared 
in terms of status prestige to Harvard or Cambridge in the ``anglo-american education 
system'' \cite{mar07}, however there are a number of other engineering schools which 
are also considered as ``Grandes \'Ecoles''. Based on our data, we find that ``Grandes \'Ecoles'' 
graduates have a success rate in their applications to permanent positions of 26.6\%, nearly three times 
that of applicants who did their undergrad studies at the university (10.6\%). 
In order to assess the significance of these results, we perform a $\chi^2$ test on these data. 
We find an overall probability  for the null-hypothesis (fortuitous occurrence 
of the factor nearly three between university and GE applicants) to be true of $p=0.0026$.
This result is reminiscent of the type of favouritism linked to the ``Grandes \'Ecoles''
and part of a general elite reproduction scheme that was identified already 
decades ago by  P. Bourdieu and is described in details in his book 
``The state nobility''\cite{bou98}. 
In this work, Bourdieu studies what he calls ``School mediated forms of [elite] reproduction''
(p 285) to access the ``field of power'' (p 267) which includes the highest positions in academia (e.g.
university professors). He defines this mechanism using a comparison with the forms of cooptation that exist 
in family-run business : ``In the [School mediated mode of reproduction], the academic title becomes a 
genuine entry pass : the school, in the form of the Grande Ecole - and the corps, a social group that 
the school produces [...] take the place of the family and family ties, with the cooptation of 
classmates based on school and corps solidarity, taking over the role played by nepotism and marital ties in business''.
The results presented here suggest that somehow the mechanisms described by
Bourdieu are at play today in the field of French astronomy. In the rest of this
paper we will refer to this specific mechanism simply as ``elitism''.

\begin{figure}
\centerline{\includegraphics[height = 6cm, keepaspectratio]{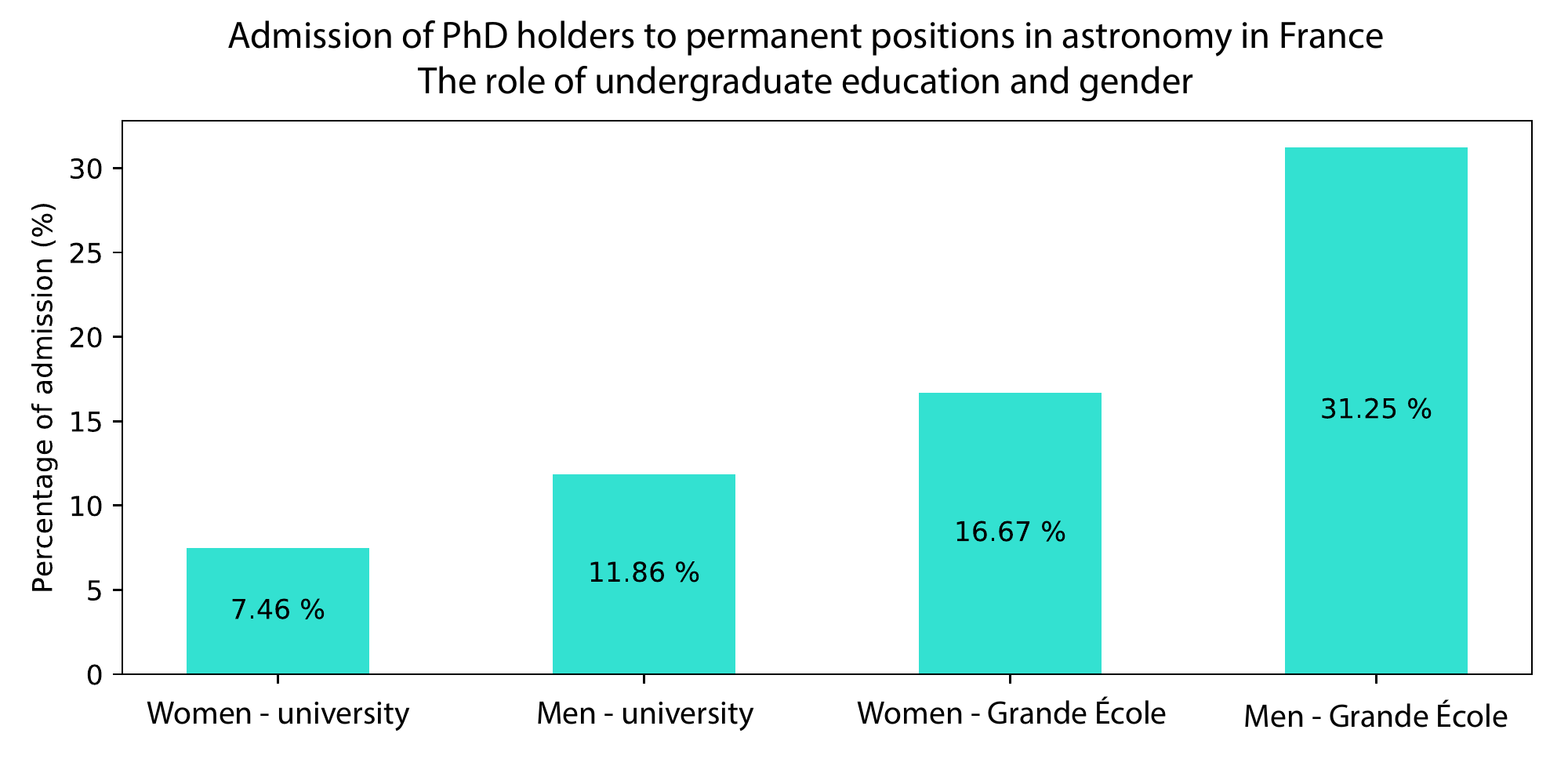}}
\caption{ Success rates of four different classes of applicants to permanent positions in astronomy in France.
\label{Fig_elite}}
\end{figure}

In Fig.~\ref{Fig_admission-rate} we now compare admission rates as a function of gender. 
It can be seen that, for all types of positions, women have 
a lower success rate then men. Overall, the success rate for men is twice that of the 
success rate for women with a significance of $p=0.06$. Although this is highly suggestive
of a gender bias, it appears to be less important than elitism. 
A question one can thus ask is whether the observed gender bias is 
somehow tied to the strong form of elitism we have identified. 
Indeed, female students are significantly under-represented in ``Grandes \'Ecoles'' :
for instance, there are 16\% and 17\% women at \'Ecole Polytechnique and 
\'Ecole Normale Superieure, respectively \cite{blan16}, whereas 
at the university, there are, according to the ministry of higher education, 52\% women 
(more specifically, for the fields of science: 28\% in ``fundamental science an applications'', 
and 60\% in ``natural science and biology''). 
Hence, if recruiting committees in French astronomy tend to favour applicants from 
``Grandes \'Ecoles'', as is the case, this could mechanically reduce the number of 
women selected at the hiring stage. 
To untangle the effects of gender and elitism, 
we define four classes of applicants: women from the university, men from the university,
women from ``Grandes \'Ecoles'', men from ``Grandes \'Ecoles'', and present their success rates
at permanent job applications in Fig.~\ref{Fig_elite}. 
This figure shows that women from ``Grandes \'Ecoles'' 
still have two times less chances of success than men from the same schools. 
Hence, the observed elitism needs to be supplemented 
by a ``genuine'' gender bias at hiring to explain the observations in Figs.
1 and 2. It is also striking that the difference in success rate between men and women from 
``Grandes \'Ecoles'' is larger than the difference in success rate of male and female 
university graduates. In a way, it seems that prestige tends to amplify at the same time 
the absolute success rate and the gender inequality.

\section{Interiorization of gender biases by women}

\begin{figure}
\centerline{\includegraphics[height = 8cm, keepaspectratio]{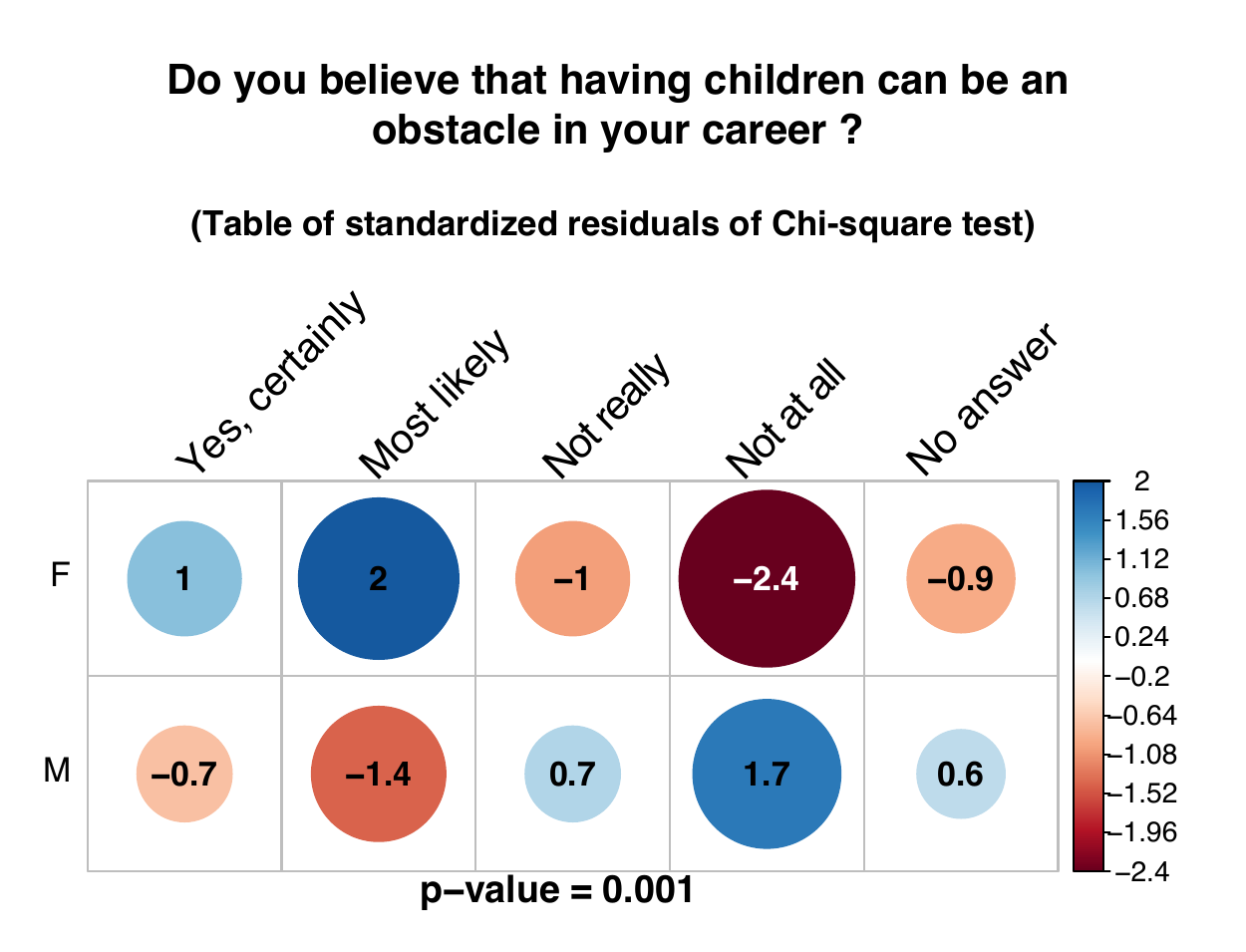}}
\caption{ Results to the question ``Do you believe that having children can be an 
obstacle in your career ?'' as a function of gender, shown as a table of standardized residuals.
These are the difference between the observed count and 
expected count in the chi-square hypothesis testing, normalized to the standard
deviation $\sigma$ of the expected count. Positive (negative) numbers $z$ 
indicate that an answer is over(under)-represented in a population at $z\times~\sigma$.
More precisely, these numbers are $Z$-scores, with $|z|>1.7$ corresponding to a probability 
$p<0.045$ in a one-tailed normal distribution. The indicated $p$-value below the table concerns 
the results of the overall $\chi^2$-test of variables ``gender'' vs ``are children an obstacle''.\label{Fig_enf-frein}}
\end{figure}

We now  investigate how such biases are interiorized (or not) by women in the 
perception of their careers. In particular, we look at their perception 
of the compatibility of pursuing a career in academia with maternity. 
We have asked in the survey if respondents  ``believe that having children can 
be an obstacle in their careers?''with 5 possible answers : ``yes certainly'', 
``yes most likely'', ``not really'', ``not at all'', ``do not want to answer''. 
In Fig.~\ref{Fig_enf-frein} we present the results to this question as a function
of gender. We find that the answer ``not at all'' is under-represented amongst 
women while the answer  ``most likely'' is over-represented. The precise opposite trend 
is observed for men, although with a smaller significance. The overall statistical
significance of these trends is however high, with $p=0.001$. We can conclude 
that women fear more the impact of having children on their careers than men. 
This feeling translates into a real life inequality : in our sample, 27.3\% of men have children
against 15.6\% for women (note that the age distributions in our sample
for men and women are similar). This shows that gender equity in terms of familial
achievements amongst astronomers is elusive, as is observed more generally in
academia \cite{mas04}. An additional interesting and perhaps non-intuitive result from our survey is 
that those who have children fear less the effect of children on their career than those 
who do not have children. 

\section{Conclusion}

In conclusion, this study suggests that at least two forms 
of biases, namely elitism and gender bias exist a the 
hiring stage in French astronomy.    
Women declare having more difficulties to conciliate  
maternity and career. As a consequence, 
fewer women in astronomy in France have children than men. 
This could be either because 
they fear that children have a negative impact on their careers, and/or
because those who do have children are somehow ``forced'' to leave
academia. National committees (such as the CNRS) have acknowledged this issue for 
several years now (B. Mosser, private communication), hence one can hope that significant 
changes will be seen in the next decade.


\section{Box: French academic system in the field of astronomy}


There are three types of permanent positions in French academia: researcher (charg\'e' de recherche), 
assistant professor (maitre de conf\'erence) and deputy astronomer (astronome adjoint).
Researchers are hired by the CNRS and do research only. Assistant professors are hired in 
universities, and teach 196 hours per year in addition to their research. Deputy astronomers 
are hired in ``observatories'' by universities and teach 
66 hours per year. Senior declinations of these positions are "Director of research", "Astronomer"
and "Professor". While it is technically possible to access to these position 
directly from a non-permanent position, this is in practice very rarely the case, and most often
these position are accessed to through career progression (e.g. from CNRS researcher to director of research). 
Researcher and dep. astronomer positions are attributed on a yearly basis through national competitive examinations
by a committee whose members are elected by the community and named by representative of 
the CNRS and ministry of research and higher education. Ass. prof. positions 
at universities are attributed locally by a jury composed of internal and external members.




\bibliographystyle{naturemag}
\bibliography{sample}





\end{document}